\documentclass[aps,prl,twocolumn,letterpaper,showpacs,10pt]{revtex4-1}
\usepackage[utf8]{inputenc}
\usepackage{graphicx,amsmath,mathrsfs}
\usepackage{amssymb}
\usepackage{amsthm}
\usepackage{bm}

\def\bc{\begin{center}}
\def\ec{\end{center}}

\newcommand{\bs}[1]{\boldsymbol{#1}}

\renewcommand {\ec}{\eta_{\gamma}}

\begin{document}
\title{Fluctuation-induced Topological Quantum Phase Transitions in Quantum Spin Hall and Quantum Anomalous Hall Insulators}
\author{Jan Carl Budich${}^{1,2}$}
\author{Ronny Thomale${}^1$}
\author{Gang Li${}^2$}
\author{Manuel Laubach${}^2$}
\author{Shou-Cheng Zhang${}^1$}
\affiliation{${}^1$Department of Physics, Stanford University, Stanford, CA 94305, USA}
\affiliation{${}^2$Institute for Theoretical Physics and Astrophysics,
University of W$\ddot{u}$rzburg, 97074 W$\ddot{u}$rzburg, Germany}
\date{\today}
\begin{abstract}
We investigate the role of quantum fluctuations in topological quantum phase transitions of quantum spin Hall insulators and quantum anomalous Hall insulators. Employing the variational cluster approximation to obtain the single-particle Green's function of the interacting many-body system,  we characterize different phases by direct calculation of the recently proposed topological order parameter for interacting systems. We pinpoint the influence of quantum fluctuations on the quantum spin Hall to Mott insulator transition in several models. Furthermore, we propose a general mechanism by which a topological quantum phase transition can be driven by the divergence of the self energy induced by interactions.
\end{abstract}
\maketitle

{\it Introduction.} Initiated by the theoretical prediction \cite{BHZ} and the experimental discovery \cite{konig2007} of the quantum spin Hall (QSH) state featuring a bulk insulating time reversal symmetry (TRS) preserving phase with topologically protected gapless helical edge modes~\cite{KMa,KMb,bernevig2006a}, the field of topological insulators (TIs) has been rapidly evolving in recent years \cite{HasanKane, xlreview}. Whereas non-interacting topological band insulators are well studied theoretically by now, interaction effects substantially complicate the picture. One focus of ongoing research hence is to develop a deeper understanding of the influence of interactions on topological quantum states of matter.

Along these lines, the topological Mott insulator~\cite{TMI} is an example where the non-interacting band structure is topologically trivial or metallic, while the interactions drive the system to the topological insulator phase. Recently, realistic materials have been predicted which could realize such a phase~\cite{zhang2011}. In an interacting system, one can not use topological invariants based on the single particle band states. Instead, a general framework based on topological field theory has been proposed, which defines topological states of matter in terms of the topological response function, {\it e.g.} the quantized Hall effect or the quantized topological magneto-electric effect~\cite{TFTTI}. These topological response functions can be expressed in terms of the single particle Green's function of an interacting system, and can be defined as the topological order parameters (TOP) of the system~\cite{TopologicalOrderParameter,WangQiZhangInversion,WangGeneralTOP}. The concept of the TOP offers a 
general framework to investigate interacting topological insulators, and its recently discovered simplification~\cite{WangQiZhangInversion,WangGeneralTOP} enables efficient practical calculations.
Recast in terms of the self energy $\Sigma(\bs{k},\omega)$ entering the single-particle Green's function, there are both static terms that renormalize the bandstructure of the non-interacting system as well as contributions from dynamic fluctuations which affect the pole structure of the self energy. Topological quantum phase transitions~\cite{BHZ} (TQPT) which are due to a critical Dirac point in the renormalized band structure can be conveniently encompassed by various mean field approaches. For example, the breakdown of the QSH phase in the presence of strong Hubbard interactions can be predicted in terms of static mean field theory \cite{RachelHur2010}. Clear signatures of interaction-induced dynamic fluctuations, however, are hard to identify unambiguously. In turn, as dynamic fluctuations represent genuine interaction effects, they are of crucial importance as to further understanding interacting topological phases~\cite{oshi}.

{\it Main results.} In this work, we demonstrate the importance of interaction-driven local dynamic fluctuations for TQPTs in two dimensional TIs as well as in quantum anomalous Hall (QAH) insulators~\cite{QAH,qi2005}, i.e. the lattice version of integer quantum Hall (QH) states. Most of our findings directly carry over to systems of arbitrary spatial dimension.
First, we calculate the phase diagram of the Kane-Mele-Hubbard model within the variational cluster approximation (VCA) not via probing the existence of edge modes in the spectral function, but via direct calculation of the TOP in the presence of inversion symmetry \cite{WangQiZhangInversion}. This  approach avoids the necessity of analyzing edge mode features of the spectral function, and might thus be indispensable for topological phases without edge modes~\cite{ashvininv,berniinv}. Comparing the phase boundary obtained from the full VCA calcuation with the static limit of our numerical data, we are able to quantitatively analyze the influence of fluctuations in the Kane-Mele-Hubbard model. By static limit, we mean the large frequency limit of the self-energy where dynamical fluctuations accounted for in our full VCA data have decayed and only frequency independent Hartree-Fock diagrams contribute. Our results explicate that many-body effects affect the phase transition from the topologically non-trivial to the trivial bands.
Second, while in previous literature strong interactions were mainly shown to destroy the nontrivial topology of a band structure, we show on the basis of a toy model how a trivial band insulator can be driven into a finite Chern number phase by means of local dynamic fluctuations. This transition has no effective single particle analogue and shows that interactions largely expand the scope of topological band structure phases.
\begin{figure}
\begin{minipage}{0.99\linewidth}
\includegraphics[width=\linewidth]{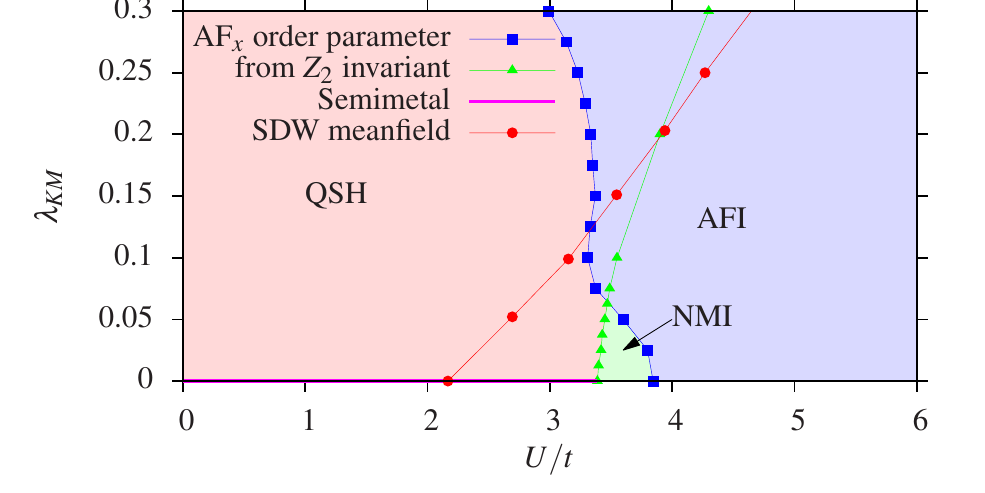}
\end{minipage}
\caption{(Color online) Phase diagram of the Kane-Mele-Hubbard model in VCA approximation obtained by direct calculation of the TOP in the presence of inversion symmetry \cite{WangQiZhangInversion}. The green triangles mark the phase boundary between the QSH and a nonmagnetic insulator (NMI) phase if a paramagnetic condition is imposed by hand in an 8-site cluster VCA calculation. The blue boxes denote the static phase boundary between the QSH and an anti-ferromagnetic (AF) insulating state in a 6-site cluster VCA calculation. The red discs show the phase boundary between the QSH and the SDW phase obtained by a self consistent  Hatree Fock calculation reported in Ref. \cite{RachelHur2010}. The light pink shaded region is our full VCA result for the QSH phase. In the light green shaded NMI region, the static result differs from the full VCA data.} 
\label{fig:phaseKM}
\end{figure}

{\it Chern numbers and $\mathbb Z_2$ invariant.} The topological information of band insulators in the presence of interactions is encoded in the homotopy of its single particle Green's function $G$~in combined frequency-momentum space \cite{VolovikBook,TFTTI,TopologicalOrderParameter}.
A perturbative expansion to lowest order of the effective action of a gauge field coupled to a gapped fermionic system  yields the Chern Simons term \cite{ChernSimons1974,Redlich1984,ZeeQuantumHallFluids}. The prefactor of this term representing the Hall conductivity of the system and its higher dimensional analogues is given by a vacuum polarization diagram the contribution of which only depends on the homotopy class of the fermionic single particle Green's function in momentum space \cite{VolovikQH3HE,Golterman1993}. This topological coupling constant takes the form
\begin{align}
\mathcal C_n = \mathcal N_n\int {\rm{d}}\omega\int{\rm{d}}^{2n}k {\rm{Tr}}\left[\left(G d G^{-1}\right)^{2n+1}\right],
\label{eqn:chernn}
\end{align}
where $\mathcal N_n$~is a normalization constant ensuring the integer-valuedness of $C_n$~and $d$~denotes the exterior derivative in combined frequency-momentum space.
This coupling constant is often times referred to as the topological charge of the system and, for a noninteracting system, reduces to the $n$-th Chern number associated with its Berry curvature \cite{TFTTI}.
Along these lines, the (4+1) dimensional time reversal symmetry (TRS) preserving analogue of the QH effect has been defined \cite{ZhangHu4DQH}.
Here, the topological charge is given by $C_2$. With the help of dimensional extension, the topological classification of 2D and 3D TIs can be reduced to analyzing the (4+1) dimensional quantum Hall phase of the respective extended system \cite{TFTTI,TopologicalOrderParameter}. The resulting TOP unifies the notion of QH and TI phases from a viewpoint of topology.

{\it TQPT due to poles of the self energy}
Whereas TQPTs in non-interacting systems are related to Dirac gap closing points in the single particle spectrum \cite{BHZ}, we encounter a much richer phenomenology in the presence of interactions \cite{Rosch2007,Gurarie2011}, as they can entail poles in the $\omega$-dependence of the self energy, i.e. zeros in the Green's function. In contrast, the Green's function of an effective static band structure is always of the form $G(k,\omega) = (i\omega- H(k))^{-1}$~implying that the frequency dependences of all effective single particle models are equivalent.  Using the relation $d^2=0$ and the fact that the combined frequency-momentum space of the physical system has no boundary, the $G\leftrightarrow G^{-1}$~symmetry of Eq. (\ref{eqn:chernn}) becomes manifest. Therefore, the poles in the self energy which are also poles of $G^{-1}$~can play a similar role as gap closing points, i.e., poles of $G$~in the non-interacting case. In general, we are not able to obtain the exact self energy of a given problem, but 
instead impose constraints on the functional form along which it becomes accesible. A commonly used local self energy approximation reads
\begin{equation}
\Sigma(\bs{k},\omega)=\Sigma_1(\bs{k})+\Sigma_2(\omega), \label{self}
\end{equation}
where $\Sigma_1$ is the previously discussed band structure renormalization and $\Sigma_2$ is the fluctuation part which is approximated to be independent of momentum $\bs{k}$. Decisive progress on the analysis of fluctuation effects versus static mean field effects has been reported recently \cite{FDWN2011,PE}.

{\it TQPT in the Kane-Mele Hubbard model.}
We now investigate the influence of local fluctuations as induced by on-site Hubbard repulsion on the QSH phase of a honeycomb lattice system.
The Hamiltonian of the Kane-Mele Hubbard model reads
\begin{align}
H= -t\sum_{<i,j>}c_i^\dag c_j+i\lambda\sum_{\ll i,j\gg}c_i^\dag \nu_{ij}\sigma_z c_j+U\sum_i n_{i\uparrow}n_{i\downarrow},
\end{align}
where $< \cdot>$~and $\ll\cdot\gg$~denote sums over nearest neighbors and next nearest neighbors respectively, $\sigma_z$~is a Pauli matrix in spin space, $c_i = (c_{i\uparrow},c_{i\downarrow})$~is a spinor, and $\nu_{ij}=\pm 1$~depends on whether the orientation of the shortest connection between the next nearest neighbors is clockwise or counter-clockwise. The variational cluster approximation (VCA) provides us with the single-particle Green's function of the interacting problem. The VCA is a quantum cluster approach which describes a given model in the thermodynamic limit via a smaller exaclty solvable reference frame (cluster) which is then associated with the infinite system within a non-perturbative variational scheme~\cite{prl-potthoff,potthoffadv}. For two spatial dimensions, we are able to keep cluster sizes that allow us to consider the leading short range correlations on an exact footing. For short-ranged interactions, the VCA hence is a rather adequate approach for topological band structures~\
cite{Yu2011, WuRachel, racheletal}.

In the absence of disorder and Rashba spin orbit interaction, besides TRS, the model possesses inversion and $S_z$~rotation symmetry. Recently, a significant simplification as to the practical calculation of the TOP of a TI in the presence of inversion symmetry has been reported \cite{WangQiZhangInversion}, which allows us to efficiently compute the phase diagram of our system by direct calculation of the TOP:
in Fig.~\ref{fig:phaseKM} we show the phase boundary of the QSH phase for various approximations. The static limit of our VCA calculation concurs with the antiferromagnetic phase boundary. Hence, in this limit, the QSH phase only breaks down in favor of antiferromagnetic order implying the spontaneous breaking of TRS. Our result is quantitively different from the results of a static spin density wave (SDW) mean field decoupling reported in Ref.~\cite{RachelHur2010} (red round dots in Fig.~\ref{fig:phaseKM}). For rather small spin orbit interaction, i.e. for small $\lambda$, the phase boundary of our full VCA calculation is not determined by the tendency of the system towards magnetic order. This indicates that at small $\lambda$~where the single particle gap becomes small, fluctuations particularly dominate the TQPT from the QSH to a nonmagnetic insulating (NMI) phase. This qualitatively different behavior as compared to the static case is cleary owing to the presence of dynamic fluctuations, i.e. to the 
nontrivial $\omega$-dependence of the self energy obtained from the VCA calculation. Quantitatively similar results for the phase boundary have been previously obtained using a complementary approach in VCA by probing the appearance of edge modes in the spectral function \cite{Yu2011}, qualitatively matching numerical microscopic calculations of the spectral properties via quantum Monte Carlo methodes which is applicable due to particle-hole symmetry in the particular model~\cite{hohenadler,congjun}.

\begin{figure}
\begin{minipage}{0.99\linewidth}
\includegraphics[width=\linewidth]{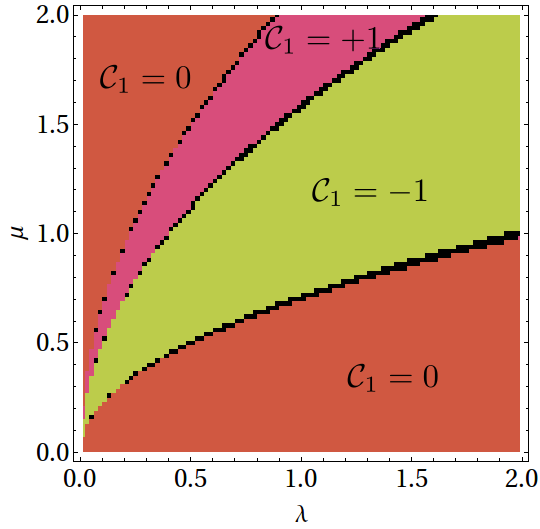}
\end{minipage}
\caption{(Color online) Phase diagram of the QAH insulator with local fluctuations. $m=-2.5$~is fixed. The pole structure of the local self energy encoded in $\lambda,\mu$~is varied. The Chern numbers $\mathcal C_1$~of the different phase domains are indicated. The black regions denote numerically unstable transition regimes.}
\label{fig:toyphase}
\end{figure}

{\it TQPT driven by the self energy.} By now, we have only discussed an example where interaction effects drive the system out of a topologically non-trivial band structure into a trivial one.
Following Ref. \cite{FDWN2011} where the self energy is assumed scalar in band space,
the static mean field part and the fluctuation-induced part of the self energy decouple and give a product of two integers yielding the TOP. Since a nontrivial $\mathbb Z_2$~invariant, i.e. an odd total TOP, can only result as a product of both these factors being odd, this rules out the possibility of a fluctuation driven transition from a trivial to a nontrivial state.
We now investigate whether such a new type of transition can still occur when we relax the diagonality constraint on the self energy.
To this end, let us consider a more general {\it matrix valued self energy} in Eq.~\ref{self} again consisting of the Hartree Fock part $\Sigma_1(\bs{k})$ and the local dynamical self energy $\Sigma_2(\omega)$. The latter can be represented as a pole expansion \cite{PE}
\begin{align}
\Sigma_2(\omega) = V^\dag (i\omega - P)^{-1}V
\end{align}
with a frequency independent Hermitian $N\times N$~matrix $P$, where $N$~is the number of poles of $\Sigma_2(\omega)$~on the imaginary axis. $V$~is an $N\times n$~matrix, where $n$~denotes the number of bands. It has been shown \cite{PE} that the TOP of the single particle Green's function $G(k,\omega)=(i\omega+\mu-h(\bs{k})-\Sigma(\bs{k},\omega))^{-1}$~with the single particle Bloch Hamiltonian $h(\bs{k})$ can be calculated by introducing an effective extended $(n+N)\times(n+N)$~single particle Hamiltonian
\begin{align}
H(\bs{k})=\begin{pmatrix}{h(\bs{k})+\Sigma_1(\bs{k})-\mu}&V^\dag\\V&P\end{pmatrix},
\end{align}
and then calculating the TOP of the single particle Green's function $\tilde G(\bs{k},\omega) = (i\omega - H(\bs{k}))^{-1}$. Note that a factorization into a non-interacting and fluctuation part is not manifest in this more general case.
We now show that a transition from trivial to nontrivial can in principle be driven by means of local fluctuations as described by a nonscalar local self energy. While our idea is very general, let us for concreteness explicitly construct a minimal toy model for such a type of transition. Consider the two band model of the QAH insulator~\cite{qi2005}
\begin{align}
h(\bs{k})= v^i(\bs{k})\sigma_i,
\label{eqn:ciham}
\end{align}
where $v^1=\sin(k_x),~v^2=\sin(k_y),~v^3=(m+\cos(k_x)+\cos(k_y))$~and $m$~is a real parameter which tunes the sign of the band gap. Assuming an interaction which brings about two poles in the self energy, we make the following ansatz for the parameters $V,P$~of the pole expansion
\begin{align}
V=\mu \sigma_x,~P=\lambda \sigma_z
\end{align}
which, for the effective extended Hamiltonian, yields
\begin{align}
H(\bs{k})=\begin{pmatrix}{h(\bs{k})}&{\mu \sigma_x}\\ {\mu\sigma_x}&{\lambda \sigma_z}\end{pmatrix}.
\label{eqn:ciextended}
\end{align}
For $m=-2.5$, the two band model is located in its trivial regime. A two dimensional phase diagram with single particle Hamiltonian defined in Eq. (\ref{eqn:ciextended}) is shown in Fig.~\ref{fig:toyphase}. We find that local fluctuations can drive a system into a topologically nontrivial phase in this model. Note that $\Sigma_1(\bs{k})=0$ in our case. Hence, exclusively $\Sigma_2(\omega)$ causes a TQPT which has no band structure analogue due to a gap closing in the two band Hamiltonian (\ref{eqn:ciham}), and is a pure fluctuation-induced phase transition that does not connect to a static mean field ansatz.

{\it Discussion.} The two fluctuation-induced transitions we describe allow us to gain a deeper understanding of the role of interactions in topological band structures. As shown above, in a fluctuation-driven TQPT, $G$~and $G^{-1}$~switch roles as compared to a transition with a critical Dirac-point in the static single particle spectrum. Here, it is the pole of the self energy and not the quasiparticle pole which determines the TQPT. Our model with non-trivial Chern number induced by interaction is an explicit example for this mechanism.

To summarize, we have identified the fingerprints of local interaction-induced dynamical fluctuations in two different kinds of TQPTs.
First, we have investigated fluctuation-driven contributions to the critical Hubbard interaction strength of the QSH-Mott transition in the Kane-Mele-Hubbard model at the level of the TOP. In our VCA calculation, the phase boundary obtained from the static limit of the single particle Green's function differs qualitatively from the full VCA data: whereas in the static limit the QSH phase breaks down only in favor of magnetic order, dynamic fluctuations retained in the full VCA result lead to a transition between the QSH and a NMI phase for smaller interactions.
Second, a proof of principle has been given for a reverse transition from a trivial band insulator to a topologically non-trivial QAH insulator, where the pole structure of the frequency dependent self energy causes a critical singularity which is substantially different from a critical Dirac point in a band inversion TQPT caused by spin orbit coupling. The fluctuation-induced effects have no band structure analogue and are a genuine signature of electronic correlations beyond mean field theory. Our arguments also carry over to three spatial dimensions, where it will be interesting to consider our findings in the framework of interaction effects for $3D$ TI's and Weyl semimetals~\cite{yongi}.

{\it Note added.} When this manuscript was finished, we became aware of complementary independent results reported in Wang et al., EPL 98, 57001 (2012).

\begin{acknowledgements}
We thank S. Rachel for important comments on applying the VCA approach to interacting electron systems with spin-orbit coupling. R.T. and M.L. thank
S. Rachel for ongoing collaborations on related topics. R.T. thanks X. Dai for comments.
J.C.B. acknowledges financial support from the DFG-JST Research Unit "Topotronics" and the hospitality of the condensed matter theory group at Stanford University. R.T. is supported by SPP 1458/1 and an SITP fellowship by Stanford University. S.C.Z. is supported by the Defense Advanced Research Projects Agency Microsystems
Technology Office, MesoDynamic Architecture Program (MESO) through the contract number
N66001-11-1-4105.
\end{acknowledgements}

\end{document}